\documentclass{bmcart}

\usepackage{amsthm,amsmath}
\RequirePackage{hyperref}
\usepackage[utf8]{inputenc} %unicode support
\usepackage{multirow}
\usepackage{graphicx}
\startlocaldefs
\endlocaldefs

\begin{document}

%%% Start of article front matter
\begin{frontmatter}

\begin{fmbox}
\dochead{Technical Note}

%%%%%%%%%%%%%%%%%%%%%%%%%%%%%%%%%%%%%%%%%%%%%%
%%                                          %%
%% Enter the title of your article here     %%
%%                                          %%
%%%%%%%%%%%%%%%%%%%%%%%%%%%%%%%%%%%%%%%%%%%%%%

\title{Second-generation PLINK: rising to the challenge of larger and richer datasets}

%%%%%%%%%%%%%%%%%%%%%%%%%%%%%%%%%%%%%%%%%%%%%%
%%                                          %%
%% Enter the authors here                   %%
%%                                          %%
%% Specify information, if available,       %%
%% in the form:                             %%
%%   <key>={<id1>,<id2>}                    %%
%%   <key>=                                 %%
%% Comment or delete the keys which are     %%
%% not used. Repeat \author command as much %%
%% as required.                             %%
%%                                          %%
%%%%%%%%%%%%%%%%%%%%%%%%%%%%%%%%%%%%%%%%%%%%%%

\author[
   addressref={aff1,aff2},                   % id's of addresses, e.g. {aff1,aff2}
   corref={aff1},                       % id of corresponding address, if any
%   noteref={n1},                        % id's of article notes, if any
   email={chrchang@alumni.caltech.edu}   % email address
]{\inits{CC}\fnm{Christopher C} \snm{Chang}}
\author[
   addressref={aff3},
   email={carsonc@mail.nih.gov}
]{\inits{CC}\fnm{Carson C} \snm{Chow}}
\author[
   addressref={aff2,aff4},
   email={laurent@cog-genomics.org}
]{\inits{LCAM}\fnm{Laurent CAM} \snm{Tellier}}
\author[
   addressref={aff3},
   email={vattikutis@niddk.nih.gov}
]{\inits{S}\fnm{Shashaank} \snm{Vattikuti}}
\author[
   addressref={aff5,aff6,aff7,aff8},
   email={shaun.purcell@mssm.edu}
]{\inits{SM}\fnm{Shaun M} \snm{Purcell}}
\author[
   addressref={aff3,aff9},
   email={leex2293@umn.edu}
]{\inits{JJ}\fnm{James J} \snm{Lee}}

%%%%%%%%%%%%%%%%%%%%%%%%%%%%%%%%%%%%%%%%%%%%%%
%%                                          %%
%% Enter the authors' addresses here        %%
%%                                          %%
%% Repeat \address commands as much as      %%
%% required.                                %%
%%                                          %%
%%%%%%%%%%%%%%%%%%%%%%%%%%%%%%%%%%%%%%%%%%%%%%

\address[id=aff1]{%                           % unique id
  \orgname{Complete Genomics}, % university, etc
  \street{2071 Stierlin Court},
  \postcode{94043}
  \city{Mountain View, CA},                               % city
  \cny{USA}                                    % country
}
\address[id=aff2]{%
  \orgname{BGI Cognitive Genomics Lab},
  \street{Building No. 11, Bei Shan Industrial Zone, Yantian District},
  \postcode{518083}
  \city{Shenzhen},
  \cny{China}
}
\address[id=aff3]{%
  \orgname{Mathematical Biology Section},
  \street{NIDDK/LBM, National Institutes of Health},
  \postcode{20892}
  \city{Bethesda, MD},
  \cny{USA}
}
\address[id=aff4]{%
  \orgname{Bioinformatics Centre},
  \street{University of Copenhagen},
  \postcode{2200}
  \city{Copenhagen},
  \cny{Denmark}
}
\address[id=aff5]{%
  \orgname{Stanley Center for Psychiatric Research},
  \street{Broad Institute of MIT and Harvard},
  \postcode{02142},
  \city{Cambridge, MA},
  \cny{USA}
}
\address[id=aff6]{%
  \orgname{Division of Psychiatric Genomics},
  \street{Department of Psychiatry, Icahn School of Medicine at Mount Sinai},
  \postcode{10029},
  \city{New York, NY},
  \cny{USA}
}
\address[id=aff7]{%
  \orgname{Institute for Genomics and Multiscale Biology},
  \street{Icahn School of Medicine at Mount Sinai},
  \postcode{10029},
  \city{New York, NY},
  \cny{USA}
}
\address[id=aff8]{%
  \orgname{Analytic and Translational Genetics Unit},
  \street{Psychiatric and Neurodevelopmental Genetics Unit, Massachusetts General Hospital},
  \postcode{02114},
  \city{Boston, MA},
  \cny{USA}
}
\address[id=aff9]{%
  \orgname{Department of Psychology},
  \street{University of Minnesota Twin Cities},
  \postcode{55455}
  \city{Minneapolis, MN},
  \cny{USA}
}

%%%%%%%%%%%%%%%%%%%%%%%%%%%%%%%%%%%%%%%%%%%%%%
%%                                          %%
%% Enter short notes here                   %%
%%                                          %%
%% Short notes will be after addresses      %%
%% on first page.                           %%
%%                                          %%
%%%%%%%%%%%%%%%%%%%%%%%%%%%%%%%%%%%%%%%%%%%%%%

\begin{artnotes}
%\note{Sample of title note}     % note to the article
% \note[id=n1]{Equal contributor} % note, connected to author
\end{artnotes}

\end{fmbox}% comment this for two column layout

%%%%%%%%%%%%%%%%%%%%%%%%%%%%%%%%%%%%%%%%%%%%%%
%%                                          %%
%% The Abstract begins here                 %%
%%                                          %%
%% Please refer to the Instructions for     %%
%% authors on http://www.biomedcentral.com  %%
%% and include the section headings         %%
%% accordingly for your article type.       %%
%%                                          %%
%%%%%%%%%%%%%%%%%%%%%%%%%%%%%%%%%%%%%%%%%%%%%%

\begin{abstractbox}

\begin{abstract} % abstract
\parttitle{Background}
PLINK 1 is a widely used open-source C/C++ toolset for genome-wide association
studies (GWAS) and research in population genetics. However, the steady
accumulation of data from imputation and whole-genome sequencing studies has
exposed a strong need for even faster and more scalable implementations of key
functions. In addition, GWAS and population-genetic data now frequently contain
probabilistic calls, phase information, and/or multiallelic variants, none of
which can be represented by PLINK 1's primary data format.

\parttitle{Findings}
To address these issues, we are developing a second-generation codebase for
PLINK.  The first major release from this codebase, PLINK 1.9, introduces
extensive use of bit-level parallelism, $O(\sqrt{n})$-time/constant-space
Hardy-Weinberg equilibrium and Fisher's exact tests, and many other algorithmic
improvements.  In combination, these changes accelerate most operations by 1-4
orders of magnitude, and allow the program to handle datasets too large to fit
in RAM.  This will be followed by PLINK 2.0, which will introduce (a) a new
data format capable of efficiently representing probabilities, phase, and
multiallelic variants, and (b) extensions of many functions to account for the
new types of information.

\parttitle{Conclusions}
The second-generation versions of PLINK will offer dramatic improvements in
performance and compatibility. For the first time, users without access to
high-end computing resources can perform several essential analyses of the
feature-rich and very large genetic datasets coming into use. 

\end{abstract}

%%%%%%%%%%%%%%%%%%%%%%%%%%%%%%%%%%%%%%%%%%%%%%
%%                                          %%
%% The keywords begin here                  %%
%%                                          %%
%% Put each keyword in separate \kwd{}.     %%
%%                                          %%
%%%%%%%%%%%%%%%%%%%%%%%%%%%%%%%%%%%%%%%%%%%%%%


% MSC classifications codes, if any
%\begin{keyword}[class=AMS]
%\kwd[Primary ]{}
%\kwd{}
%\kwd[; secondary ]{}
%\end{keyword}

\end{abstractbox}
%
%\end{fmbox}% uncomment this for twcolumn layout

\end{frontmatter}

%%%%%%%%%%%%%%%%%%%%%%%%%%%%%%%%%%%%%%%%%%%%%%
%%                                          %%
%% The Main Body begins here                %%
%%                                          %%
%% Please refer to the instructions for     %%
%% authors on:                              %%
%% http://www.biomedcentral.com/info/authors%%
%% and include the section headings         %%
%% accordingly for your article type.       %%
%%                                          %%
%% See the Results and Discussion section   %%
%% for details on how to create sub-sections%%
%%                                          %%
%% use \cite{...} to cite references        %%
%%  \cite{koon} and                         %%
%%  \cite{oreg,khar,zvai,xjon,schn,pond}    %%
%%  \nocite{smith,marg,hunn,advi,koha,mouse}%%
%%                                          %%
%%%%%%%%%%%%%%%%%%%%%%%%%%%%%%%%%%%%%%%%%%%%%%

%%%%%%%%%%%%%%%%%%%%%%%%% start of article main body
% <put your article body there>

%%%%%%%%%%%%%%%%
%% Background %%
%%
\section*{Findings}
Because of its broad functionality and efficient binary file format, PLINK is
widely employed in data-processing pipelines set up for gene-trait mapping and
population-genetic studies.  The five years since the final first-generation
update (v1.07), however, have witnessed the introduction of new algorithms and
analytical approaches, the growth in size of typical datasets, and wide
deployment of heavily multicore processors.

In response, we have developed PLINK 1.9, a comprehensive performance, scaling,
and usability update.  Its speed improvements are the most notable: our data
indicate that speedups frequently exceed two, and sometimes even three, orders
of magnitude for several commonly used operations.  Its core functional domains
are unchanged from that of its predecessor (data management, summary
statistics, population stratification, association analysis,
identity-by-descent estimation \cite{purcell2007}), and it is usable as a
drop-in replacement in most cases, requiring no changes to existing scripts.
To support easier interoperation with newer software like BEAGLE 4
\cite{browning2013}, IMPUTE2 \cite{howie2009}, GATK \cite{mckenna2010},
VCFtools \cite{danecek2011}, BCFtools \cite{li2009}, and GCTA \cite{yang2011},
features such as the import/export of VCF and Oxford-format files and an
efficient cross-platform genomic relationship matrix (GRM) calculator have been
introduced.  Most pipelines currently employing PLINK can expect to benefit
from upgrading.

A major problem remains: PLINK's core file format can only represent unphased,
biallelic data.  We are developing a second update, PLINK 2.0, to address this.

\subsection*{Improvements in PLINK 1.9}

\subsubsection*{Bitwise parallelism}
Modern x86 processors are designed to operate on data in (usually 64-bit)
machine word or ($\geq $ 128-bit) vector chunks.  The PLINK 1 binary file
format supports this exceptionally well: its packed 2-bit data elements can,
with the use of bit arithmetic, easily be processed 32 or 64 at a time.
However, most existing programs fail to exploit opportunities for bitwise
parallelism; instead their loops painstakingly extract and operate on a single
data element at a time.  Replacement of these loops with bit-parallel logic is,
by itself, enough to speed up numerous operations by more than one order of
magnitude.

For example, the old identity-by-state calculation proceeded roughly as
follows:

\vspace*{\baselineskip}

For every sample pair $(i, j)$:
\begin{itemize}
  \item[] For every marker $k$:
  \begin{enumerate}
    \item If either $i_k$ or $j_k$ is a missing call, skip
    \item If $i_k=j_k$, increment IBS2 count
    \item otherwise, if both bits differ, increment IBS0 count
    \item otherwise, increment IBS1
  \end{enumerate}
\end{itemize}

\vspace*{\baselineskip}

We replaced this with:

\vspace*{\baselineskip}

For every sample pair $(i, j)$:
\begin{itemize}
  \item[] For every 960-marker block $K$:
  \begin{enumerate}
    \item Evaluate $i_K$ XOR $j_K$
    \item Mask out markers with missing calls
    \item Count number of set bits
  \end{enumerate}
\end{itemize}

\vspace*{\baselineskip}

Refer to Additional file 1 for a detailed walkthrough.  Our timing data (see
``Performance comparisons'' below) indicate that PLINK 1.9 takes less than
twice as long to handle a 960-marker block as PLINK 1.07 takes to handle a
single marker.

\subsubsection*{Bit population count}

The last step above---bit ``population count''---merits further discussion.
Post-2008 x86 processors support a specialized instruction that directly
evaluates this quantity.  However, thanks to 50 years of work on the problem,
algorithms exist which evaluate bit population count nearly as quickly as the
hardware instruction, while sticking to universally available operations.
Since PLINK is still used on some older machines, we took one such algorithm
(previously discussed and refined by Dalke, Harley, Lauradoux, Mathisen, and
Walisch \cite{dalkeurl}), and developed an improved SSE2-based implementation.
(Note that SSE2 vector instructions are supported by even the oldest x86-64
processors.)

The applications of bit population count extend further than might be obvious
at first glance.  As an example, consider computation of the correlation
coefficient $r$ between a pair of markers, where some data may be missing.
Letting $x$ and $y$ denote the markers, $i\in S$ denote sample indices, define
$x_i$ and $y_i$ to be $-1$ when the corresponding genotype call is homozygous
minor, 0 when the corresponding call is heterozygous or missing, and $+1$ when
the corresponding call is homozygous major.  Also define $S_{xy}$ to be the
subset of $S$ for which $x$ and $y$ do not have missing calls,
$\overline{x} := |S_{xy}|^{-1}\sum_{i\in S_{xy}}x_i$ (similarly for
$\overline{y}$), and $\overline{x^2} := |S_{xy}|^{-1}\sum_{i\in S_{xy}}x_i^2$
(similarly for $\overline{y^2}$).  ($|\cdot |$ denotes set size.)  The
correlation coefficient can then be expressed as

\begin{eqnarray}
r & = & \frac{|S_{xy}|^{-1}\sum_{i\in S_{xy}}(x_i - \overline{x})(y_i - \overline{y})}{\sqrt{(\overline{x^2} - \overline{x}^2)(\overline{y^2} - \overline{y}^2)}} \nonumber \\
& = & \frac{|S_{xy}|^{-1}\sum_{i\in S}x_iy_i - \overline{x}\cdot \overline{y}}{\sqrt{(\overline{x^2} - \overline{x}^2)(\overline{y^2} - \overline{y}^2)}}
\nonumber
\end{eqnarray}

Given PLINK 1 binary data, $|S_{xy}|$, $\overline{x}$, $\overline{y}$,
$\overline{x^2}$, and $\overline{y^2}$ can easily be expressed in terms of bit
population counts.  (When no missing calls are present, these values can be
precomputed since they do not vary between marker pairs; but in the general
case, it is necessary to recalculate them all in the inner loop.)  The dot
product $\sum_{i\in S}x_iy_i$ is trickier; to evaluate it, we preprocess the
data so that the genotype bit vectors $G_x$ and $G_y$ encode homozygote minor
calls as \texttt{00}$_2$, heterozygote and missing calls as \texttt{01}$_2$,
and homozygote major calls as \texttt{10}$_2$, and then proceed as follows:

\vspace*{\baselineskip}

\begin{enumerate}
  \item Set \texttt{$G_z$ := ($G_x$ OR $G_y$) AND 01010101...$_2$}
  \item Evaluate
  \begin{itemize}
    \item[]
      \texttt{popcount2((($G_x$ XOR $G_y$) AND (10101010...$_2$ - $G_z$)) OR $G_z$)},
  \end{itemize}
  where \texttt{popcount2()} sums 2-bit quantities instead of counting set
  bits.  (This is actually cheaper than regular population count; the first
  step of software \texttt{popcount()} is reduction to a \texttt{popcount2()}
  problem.)
  \item Subtract the latter quantity from $|S|$.
\end{enumerate}

\vspace*{\baselineskip}

The key insight behind this implementation is that each $x_iy_i$ term is in
$\{-1, 0, 1\}$, and can still be represented in 2 bits.  (This is not strictly
necessary for bitwise parallel processing---the partial sum lookup algorithm
discussed later handles 3-bit outputs by padding the raw input data to 3 bits
per genotype call---but it allows for unusually high efficiency.)  The exact
sequence of operations that we chose to evaluate the dot-product terms in a
bitwise parallel fashion is somewhat arbitrary.

See \texttt{popcount\_longs()} in \texttt{plink\_common.c} for our primary bit
population count function, and \texttt{plink\_ld.c} for several correlation
coefficient evaluation functions.

\subsubsection*{Multicore and cluster parallelism}

Modern x86 processors also contain increasing numbers of cores, and
computational workloads in genetic studies tend to contain large
``embarrassingly parallel'' steps which can easily exploit additional cores.
Therefore, PLINK 1.9 autodetects the number of cores present in the machine it
is running on, and many of its heavy-duty operations default to employing
roughly that number of threads.  (This behavior can be manually controlled with
the \texttt{--threads} flag.)  Most of PLINK 1.9's multithreaded computations
use a simple set of cross-platform C functions and macros, which compile to
\texttt{pthread} library idioms on Linux and OS X, and OS-specific idioms like
\texttt{\_beginthreadex()} on Windows.

PLINK 1.9 also contains improved support for distributed computation: the
\texttt{--parallel} flag makes it easy to split large matrix computations
across a cluster.

One major computational resource remains unexploited: graphics processing
units.  We have made development of GPU-specific code a low priority since
their installed base is much smaller than that of multicore processors, and the
speedup factor over well-written multithreaded code running on similar-cost,
less specialized hardware is usually less than 10x \cite{lee2010}.  However, we
do plan to build out GPU support for the heaviest-duty computations after most
of our other PLINK 2 development goals are achieved.

\subsubsection*{Memory efficiency}

To make it possible for PLINK 1.9 to handle the huge datasets which benefit the
most from these speed improvements, the program core no longer keeps the main
genomic data matrix in memory; instead, most of its functions only load data
for a single marker, or a small window of markers, at a time.  Sample $\times$
sample matrix computations still normally require additional memory
proportional to the square of the sample size, but \texttt{--parallel} gets
around this:

\vspace*{\baselineskip}

\texttt{plink --bfile} [fileset name] \texttt{--make-grm-bin --parallel 1 40}

\texttt{plink --bfile} [fileset name] \texttt{--make-grm-bin --parallel 2 40}

...

\texttt{plink --bfile} [fileset name] \texttt{--make-grm-bin --parallel 40 40}

\texttt{cat plink.grm.bin.1} ... \texttt{plink.grm.bin.40 > plink.grm.bin}

\texttt{cat plink.grm.N.bin.1} ... \texttt{plink.grm.N.bin.40 > plink.grm.N.bin}

\vspace*{\baselineskip}

\noindent calculates 1/40th of the genomic relationship matrix per run, with
correspondingly reduced memory requirements.

\subsubsection*{Other noteworthy algorithms}

\paragraph*{Partial sum lookup}

Each entry of a weighted distance matrix is a sum of per-marker terms.  Given
PLINK 1 binary data, for any specific marker, there are at most seven distinct
cases:

\vspace*{\baselineskip}

\begin{enumerate}
\item Both genotypes are homozygous for the major allele.
\item One is homozygous major, and the other is heterozygous.
\item One is homozygous major, and the other is homozygous minor.
\item Both are heterozygous.
\item One is heterozygous, and the other is homozygous minor.
\item Both are homozygous minor.
\item At least one genotype is missing.
\end{enumerate}

\vspace*{\baselineskip}

For example, the GCTA genomic relationship matrix is defined by the following
per-marker increments (where $q$ is the minor allele frequency):

\vspace*{\baselineskip}

\begin{enumerate}
\item $\frac{(2-2q)(2-2q)}{2q(1-q)}$
\item $\frac{(2-2q)(1-2q)}{2q(1-q)}$
\item $\frac{(2-2q)(0-2q)}{2q(1-q)}$
\item $\frac{(1-2q)(1-2q)}{2q(1-q)}$
\item $\frac{(1-2q)(0-2q)}{2q(1-q)}$
\item $\frac{(0-2q)(0-2q)}{2q(1-q)}$
\item $0$ (subtract $1$ from the final denominator instead, in another loop)
\end{enumerate}

\vspace*{\baselineskip}

This suggests the following matrix calculation algorithm, as a first draft:

\vspace*{\baselineskip}

\begin{enumerate}
\item Initialize all distance/relationship partial sums to zero.
\item For each marker, calculate and save the seven possible increments in a
lookup table, and then refer to the table when updating partial sums.  This
replaces several floating point adds/multiplies in the inner loop with a single
addition operation.
\end{enumerate}

\vspace*{\baselineskip}

We can substantially improve on this by handling multiple markers at a time.
Since seven cases can be distinguished by three bits, we can compose a sequence
of operations which maps a pair of padded 2-bit genotypes to seven different
3-bit values in the appropriate manner.  On 64-bit machines, 20 3-bit values
can be packed into a machine word (for example, let bits 0-2 describe the
relation at marker \#0, bits 3-5 describe the relation at marker \#1, etc., all
the way up to bits 57-59 describing the relation at marker \#19), so this
representation lets us instruct the processor to act on 20 markers
simultaneously.

Then, we need to perform the update
\begin{equation*}
A_{jk} := A_{jk} + f_0(x_0) + f_1(x_1) + \ldots + f_{19}(x_{19})
\end{equation*}
where the $x_i$'s are bit trios, and the $f_i$'s map them to increments.  This
could be done with 20 table lookups and floating point addition operations.
Or, the update could be restructured as
\begin{equation*}
A_{jk} := A_{jk} + f_{\{0-4\}}(x_{\{0-4\}}) + \ldots + f_{\{15-19\}}(x_{\{15-19\}})
\end{equation*}
where $x_{\{0-4\}}$ denotes the lowest-order \textbf{15} bits, and
$f_{\{0-4\}}$ maps them directly to
$f_0(x_0) + f_1(x_1) + f_2(x_2) + f_3(x_3) + f_4(x_4)$; similarly for
$f_{\{5-9\}}$, $f_{\{10-14\}}$, and $f_{\{15-19\}}$.  In exchange for some
precomputation (four tables with $2^{15}$ entries each; total size 1 MB, which
is not onerous for modern L2/L3 caches), this restructuring licenses the use of
four table lookups and adds per update instead of twenty.  See
\texttt{fill\_weights\_r()} and \texttt{incr\_dists\_r()} in
\texttt{plink\_calc.c} for source code.

\paragraph*{Hardy-Weinberg and Fisher's exact tests}

PLINK 1.0 used Wigginton et al.'s SNP-HWE algorithm \cite{wigginton2005} to
test for Hardy-Weinberg equilibrium, and Mehta et al.'s FEXACT network
algorithm \cite{mehta1986} \cite{clarkson1993} for Fisher's exact test on
$2 \times 2$ and $2 \times 3$ tables.

SNP-HWE exploits the fact that, while the absolute likelihood of a contingency
table involves large factorials which are fairly expensive to evaluate, the
ratios between its likelihood and that of adjacent tables are simple since the
factorials almost entirely cancel out.  While studying the software, we made
two additional observations:

\vspace*{\baselineskip}

\begin{enumerate}
\item Its size-$O(n)$ memory allocation (where $n$ is the sum of all
  contingency table entries) could be avoided by reordering the
  calculation; it is only necessary to track a few partial sums.
\item Since likelihoods decay super-geometrically as one moves away from the
  most probable table, only $O(\sqrt{n})$ of the likelihoods can meaningfully
  impact the partial sums; the sum of the remaining terms is too small to
  consistently affect even the 10th significant digit in the final p-value.  By
  terminating the calculation when all the partial sums stop changing (due to
  the newest term being too tiny to be tracked by IEEE-754 double-precision
  numbers), computational complexity is reduced from $O(n)$ to $O(\sqrt{n})$
  with no loss of precision.  See Figure 1 for an example.
\end{enumerate}

\vspace*{\baselineskip}

Fisher's exact test for $2 \times 2$ tables has the same mathematical
structure, so it was straightforward to modify the early-termination SNP-HWE
algorithm to handle it.  The $2 \times 3$ case is more complicated, but retains
the property that only $O(\sqrt{\mbox{\# of tables}})$ relative likelihoods
need to be evaluated, so we were able to develop a function to handle it in
$O(n)$ time.  Our timing data indicate that our new functions represent very
large improvements over both FEXACT and Requena et al.'s updates
\cite{requena2006} to the network algorithm.

Standalone source code for early-termination SNP-HWE and Fisher's
$2 \times 2$/$2 \times 3$ exact test is posted at \cite{changfisherurl}. (Due
to recent calls for use of mid-$p$ adjustments in biostatistics
\cite{lydersen2009} \cite{graffelman2013}, all of these functions have mid-$p$
modes, and PLINK 1.9 exposes them.)  We are preparing another paper which
discusses these algorithms in more detail, with attention to numerical
stability and a full explanation of how the Fisher's exact test algorithm
extends to larger tables.

\paragraph*{Haplotype block estimation}

PLINK 1.0's \texttt{--blocks} command implements Gabriel et al.'s
\cite{gabriel2002} confidence interval-based method of estimating haplotype
blocks.  (More precisely, it is a restricted port of Haploview's
\cite{barrett2005} implementation of the method.)  Briefly, the method involves
using 90\% confidence intervals for $D^\prime$ (as defined by Wall and
Pritchard \cite{wall2003}) to classify pairs of variants as ``strong LD'',
``strong evidence for historical recombination'', or ``inconclusive''; then,
contiguous groups of variants where ``strong LD'' pairs outnumber
``recombination'' pairs by more than 19 to 1 are greedily selected, starting
with the longest base-pair spans.

PLINK 1.9 accelerates this in several ways:

\begin{itemize}
\item Determination of the initial diplotype frequency and $D^\prime$ point
estimates has been streamlined.  We use the analytic solution to Hill's
diplotype frequency cubic equation \cite{gaunt2007}, and only compute log
likelihoods when multiple solutions to the equation are in the valid range.
\item 90\% confidence intervals were originally estimated by computing relative
likelihoods at 101 points (corresponding to $D^\prime =0, D^\prime =0.01,
\ldots, D^\prime= 1$) and checking where the resulting cumulative distribution
function crossed 5\% and 95\%.  However, the likelihood function rarely has
more than one extreme point in $(0, 1)$ (and the full solution to the cubic
equation reveals the presence of additional extrema); it is usually possible to
exploit this property to establish good bounds on key cdf values after
evaluating just a few likelihoods.  In particular, many confidence intervals
can be classified as ``recombination'' after inspection of just two of the 101
points; see Figure 2.
\item Instead of saving the classification of every variant pair and looking up
the resulting massive table at a later point, we just update a small number of
``strong LD pairs within last $k$ variants'' and ``recombination pairs within
last $k$ variants'' counts while processing the data sequentially, saving only
final haploblock candidates.  This reduces the amount of time spent looking up
out-of-cache memory, and also allows much larger datasets to be processed.
\item Since ``strong LD'' pairs must outnumber ``recombination'' pairs by 19 to
1, it does not take many ``recombination'' pairs in a window before one can
prove no haploblock can contain that window.  When this bound is crossed, we
take the opportunity to entirely skip classification of many pairs of variants.
\end{itemize}

Most of these ideas are implemented in \texttt{haploview\_blocks\_classify()}
and \texttt{haploview\_blocks()} in \texttt{plink\_ld.c}.  The last two
optimizations were previously implemented in Taliun's ``LDExplorer'' R package
\cite{taliun2014}.

\paragraph*{Coordinate-descent LASSO}

PLINK 1.9 includes a basic coordinate-descent LASSO implementation
\cite{friedman2007} (\texttt{--lasso}), which can be useful for phenotypic
prediction and related applications.  See Vattikuti et al. \cite{vattikuti2014}
for discussion of its theoretical properties.

\subsubsection*{Newly integrated third-party software}

\paragraph*{PLINK 1.0 commands}

Many teams have significantly improved upon PLINK 1.0's implementations of
various commands and made their work open source.  In several cases, their
innovations have been integrated into PLINK 1.9; examples include

\begin{itemize}
\item Pahl et al.'s PERMORY algorithm for fast permutation testing
  \cite{steiss2012},
\item Wan et al.'s BOOST software for fast epistasis testing \cite{wan2010},
\item Ueki, Cordell, and Howey's \texttt{--fast-epistasis} variance correction
  and joint-effects test \cite{ueki2012} \cite{howeyurl}, and
\item Pascal Pons's winning submission to the GWAS Speedup logistic regression
  crowdsourcing contest \cite{ponsurl}.  (The contest was designed by Po-Ru
  Loh, run by Babbage Analytics \& Innovation and TopCoder, and subsequent
  analysis and code preparation were performed by Andrew Hill, Ragu Bharadwaj,
  and Scott Jelinsky.  A manuscript is in preparation by these authors and Iain
  Kilty, Kevin Boudreau, Karim Lakhani and Eva Guinan.)
\end{itemize}

In all such cases, PLINK's citation instructions direct users of the affected
functions to cite the original work.

\paragraph*{Multithreaded gzip}

For many purposes, compressed text files strike a good balance between ease of
interpretation, loading speed, and resource consumption.  However, the
computational cost of generating them is fairly high; it is not uncommon for
data compression to take longer than all other operations combined.  To make a
dent in this bottleneck, we have written a simple multithreaded compression
library function based on Mark Adler's excellent \texttt{pigz} program
\cite{adlerurl}, and routed most of PLINK 1.9's gzipping through it.  See
\texttt{parallel\_compress()} in \texttt{pigz.c} for details.

\subsubsection*{Convenience features}

\paragraph*{Import and export of VCF- and Oxford-formatted data}

PLINK 1.9 can import data from VCF/BCF2 (\texttt{--vcf}, \texttt{--bcf}) and
Oxford-format (\texttt{--data}, \texttt{--bgen}) files.  However, since it
cannot handle probabilistic calls, phase information, or variants with more
than two alleles, the import process is currently quite lossy.  Specifically,

\begin{itemize}
\item With Oxford-format files, genotype likelihoods smaller than 0.9 are
  normally treated as missing calls (and the rest are treated as hard calls);
  \texttt{--hard-call-threshold} can be used to change the threshold, or
  request independent pseudorandom calls based on the likelihoods in the file.
\item Phase is discarded.
\item By default, when a VCF variant has more than one alternate allele, only
  the most common alternate is retained (all other alternate calls are
  converted to missing).  \texttt{--biallelic-only} can be used to skip
  variants with multiple alternate alleles.
\end{itemize}

Export to these formats is also possible, via \texttt{--recode vcf} and
\texttt{--recode oxford}.

\paragraph*{Nonstandard chromosome code support}

When the \texttt{--allow-extra-chr} or \texttt{--aec} flag is used, PLINK 1.9
allows datasets to contain unplaced contigs or other arbitrary chromosome
names, and most commands will handle them in a reasonable manner.  Also,
arbitrary nonhuman species (with haploid or diploid genomes) can now be
specified with \texttt{--chr-set}.

\paragraph*{Command-line help}

To improve the experience of using PLINK interactively, we have expanded the
\texttt{--help} flag's functionality.  When invoked with no parameters, it now
prints an entire mini-manual.  Given keyword(s), it instead searches for and
prints mini-manual entries associated with those keyword(s), and handles
misspelled keywords and keyword prefixes in a reasonable manner.

\subsection*{A comment on within-family analysis}

Most of our discussion has addressed computational issues. There is one methodological issue, however, that deserves a brief comment. The online documentation of PLINK 1.07 weighed the pros and cons of its permutation procedure for within-family analysis of quantitative traits (QFAM) with respect to the standard quantitative transmission disequilibrium test (QTDT). It pointed out that likelihood-based QTDT enjoyed the advantages of computational speed and increased statistical power. However, a comparison of statistical power is only meaningful if both procedures are anchored to the same Type 1 error rate with respect to the null hypothesis of no linkage with a causal variant, and Ewens et al. \cite{ewens2008} have shown that the QTDT is not robust against certain forms of confounding (population stratification). The validity of a permutation procedure such as QFAM, on the other hand, only depends on the applicability of Mendel's laws. When this nicety is combined with the vast speedup of permutation in PLINK 1.9, a given user may now decide to rate QFAM more highly relative to QTDT when considering available options for within-family analysis.

\subsection*{Performance comparisons}

In the following tables, running times are collected from seven machines
operating on three datasets.

\begin{itemize}
\item ``Mac-2'' denotes a MacBook Pro with a 2.8 Ghz Intel Core 2 Duo processor
and 4GB RAM running OS X 10.6.8.
\item ``Mac-12'' denotes a Mac Pro with two 2.93 Ghz Intel 6-core Xeon
processors and 64GB RAM running OS X 10.6.8.
\item ``Linux32-2'' denotes a machine with a 2.4 Ghz Intel Core 2 Duo E6600
processor and 1GB RAM running 32-bit Ubuntu Linux.
\item ``Linux32-8'' denotes a machine with a 3.4 Ghz Intel Core i7-3770
processor (8 cores) and 8GB RAM running 32-bit Ubuntu Linux.
\item ``Linux64-512'' denotes a machine with sixty-four AMD 8-core Opteron 6282
SE processors and 512GB RAM running 64-bit Linux.
\item ``Win32-2'' denotes a laptop with a 2.4 Ghz Intel Core i5-2430M processor
(2 cores) and 4GB RAM running 32-bit Windows 7 SP1.
\item ``Win64-2'' denotes a machine with a 2.3 Ghz Intel Celeron G1610T
processor (2 cores) and 8GB RAM running 64-bit Windows 8.
\item ``synth1'' refers to a 1000 sample, 100000 variant synthetic dataset
generated with HAPGEN2 \cite{su2011}, while ``synth1p'' refers to the same
dataset after one round of \texttt{--indep-pairwise 50 5 0.5} pruning (with
76124 markers remaining).  For case/control tests, PLINK 1.9's
\texttt{--tail-pheno 0} command was used to downcode the quantitative phenotype
to case/control.
\item ``synth2'' refers to a 4000 case, 6000 control synthetic dataset with
88025 markers on chromosomes 19-22 generated by resampling HapMap and 1000
Genomes data with simuRare \cite{xu2013} and then removing monomorphic loci.
``synth2p'' refers to the same dataset after one round of
\texttt{--indep-pairwise 700 70 0.7} pruning (with 71307 markers remaining).
\item ``1000g'' refers to the entire 1092 sample, 39637448 variant 1000 Genomes
project phase 1 dataset \cite{1000genomes}.  ``chr1'' refers to chromosome 1
from this dataset, with 3001739 variants.  ``chr1snp'' refers to chromosome 1
after removal of all non-SNPs and one round of
\texttt{--indep-pairwise 20000 2000 0.5} pruning (798703 markers remaining).
\end{itemize}

All times are in seconds.  To reduce disk-caching variance, timing runs are
preceded by ``warmup'' commands like \texttt{plink --freq}.  PLINK 1.07 was run
with the \texttt{--noweb} flag.  ``nomem'' indicates that the program ran out
of memory and there was no low-memory mode or other straightforward workaround.
A tilde indicates that runtime was extrapolated from several smaller problem
instances.

\subsubsection*{Initialization and basic I/O}

Table 1 displays execution times for \texttt{plink --freq}, one of the simplest
operations PLINK can perform.  These timings reflect fixed initialization and
I/O overhead.  (Due to the use of warmup runs, they do not include disk
latency.)

\subsubsection*{Identity-by-state matrices, complete linkage clustering}

The PLINK 1.0 \texttt{--cluster --matrix} flag combination launches an
identity-by-state matrix calculation and writes the result to disk, and then
performs complete linkage clustering on the data; when \texttt{--ppc} is added,
a pairwise population concordance constraint is applied to the clustering
process.  As discussed earlier, PLINK 1.9 employs an XOR/bit population count
algorithm which speeds up the matrix calculation by a large constant factor;
the computational complexity of the clustering algorithm has also been reduced,
from $O(n^3)$ to $O(n^2 \log n)$.  (Further improvement of clustering
complexity, to $O(n^2)$, is possible in some cases \cite{defays1977}.)

In Table 2, we compare PLINK 1.07 and PLINK 1.9 execution times under three
scenarios: IBS matrix calculation only (\texttt{--cluster --matrix --K} [sample
count - 1] in PLINK 1.07, \texttt{--distance ibs square} in PLINK 1.9), IBS
matrix + standard clustering (\texttt{--cluster --matrix} for both versions),
and IBD report generation (\texttt{--Z-genome}).

(Note that newer algorithms such as BEAGLE's fastIBD \cite{browning2011}
generate more accurate IBD estimates than PLINK \texttt{--Z-genome}.  However,
the \texttt{--Z-genome} report contains other useful information.)

\subsubsection*{Genomic relationship matrices}

GCTA's \texttt{--make-grm-bin} command (\texttt{--make-grm} in early versions)
calculates the variance-standardized genomic relationship matrix used by many
of its other commands.  The latest implementation as of this writing is very
fast, but cannot run on OS X or Windows.  PLINK 1.9 includes a cross-platform
implementation which is almost as fast and has a lighter memory requirement.
See Table 3 for timing data.  (The comparison is with GCTA v1.24 on 64-bit
Linux, and v1.02 elsewhere.)

\subsubsection*{Linkage disequilibrium-based variant pruning}

The PLINK 1.0 \texttt{--indep-pairwise} command is frequently used in
preparation for analyses which assume approximate linkage equilibrium.  In
Table 4, we compare PLINK 1.07 and PLINK 1.9 execution times for some
reasonable parameter choices.  Note that as of this writing,
\texttt{--indep-pairwise}'s implementation is single-threaded; this is why the
heavily multicore machines are not faster than the 2-core machines.  The $r^2$
threshold for ``synth2'' was chosen to make the ``synth1p'' and ``synth2p''
pruned datasets contain similar number of SNPs, so Tables 2-3 could clearly
demonstrate scaling w.r.t. sample size.

\subsubsection*{Haplotype block estimation}

Table 5 demonstrates the impact of our rewrite of \texttt{--blocks}.  Due to a
minor bug in PLINK 1.0's handling of low-MAF variants, we pruned each dataset
to contain only variants with MAF $\ge 0.05$ before running \texttt{--blocks}.
95506 markers remained in the ``synth1'' dataset, and 554549 markers remained
in ``chr1''.  A question mark indicates that the extrapolated runtime may not
be valid since we suspect Haploview or PLINK 1.0 would have run out of memory
before finishing.

\subsubsection*{Association analysis max(T) permutation tests}

PLINK 1.0's basic association analysis commands were quite flexible, but the
powerful max(T) permutation test suffered from poor performance.  PRESTO
\cite{browning2008} and PERMORY introduced major algorithmic improvements
(including bit population count) which largely solved the problem.  Table 6
shows that PLINK 1.9 successfully extends the PERMORY algorithm to the full
range of PLINK 1.0's association analyses, while making Fisher's exact test
practical to use in permutation tests.  (There is no 64-bit Windows PERMORY
build, so the comparisons on the Win64-2 machine are between 64-bit PLINK and
32-bit PERMORY.)

\subsection*{PLINK 2 design}

Despite its computational advances, we recognize that PLINK 1.9 can ultimately
still be an unsatisfactory tool for working with imputed genomic data, due to
the limitations of the PLINK 1 binary file format.  To address this, PLINK 2.0
will support a new core file format capable of representing essentially all
information emitted by modern imputation tools, and many of its functions will
be extended to account for the new types of information.

\subsubsection*{Multiple data representations}

As discussed earlier, PLINK 1 binary is inadequate in three ways: probabilities
strictly between 0 and 1 cannot be represented, phase cannot be stored, and
variants are limited to two alleles.  This can be addressed by representing
\textit{all} calls probabilistically, and introducing a few other extensions.
Unfortunately, this would make PLINK 2.0's representation of PLINK 1-format
data so inefficient that it would amount to a serious downgrade from PLINK 1.9
for many purposes.

Therefore, our new format defines several data representations, one of which is
equivalent to PLINK 1 binary, and allows different files, or even variants
within a single file, to use different representations.  To work with this,
PLINK 2 will include a translation layer which allows individual functions to
assume a specific representation is used.  As with the rest of PLINK's source
code, this translation layer will be open source and usable in other programs
under GPLv3 terms; and unlike most of the other source code, it will be
explicitly designed to be included as a standalone library.  PLINK 2 will also
be able to convert files/variants from one data representation to another,
making it practical for third-party tools lacking access to the library to
demand a specific representation.

\subsubsection*{Data compression}

PLINK 1.9 demonstrates the power of a weak form of compressive genomics
\cite{loh2012}: by using bit arithmetic to perform computation directly on
compressed genomic data, it frequently exhibits far better performance than
programs which require an explicit decompression step.  But its ``compressed
format'' is merely a tight packing which does not support the holy grail of
true sublinear analysis.

To do our part to make ``strong'' sublinear compressive genomics a reality, the
PLINK 2 file format will introduce support for ``deviations from reference''
storage of low-MAF variants.  For datasets containing many samples, this
captures much of the storage efficiency benefit of having real reference
genomes available, without the drawback of forcing all programs operating on
the data to have access to a library of references.  Thanks to PLINK 2's
translation layer and file conversion facilities, programmers will be able to
ignore this feature during initial development of a tool, and then work to
exploit it after basic functionality is in place.

We note that LD-based compression of variant groups is also possible, and
Sambo's SNPack software \cite{sambo2014} applies this to the PLINK 1 binary
format.  We do not plan to support this in PLINK 2.0 due to the additional
software complexity required to handle probabilistic and multiallelic data, but
we believe this is a promising avenue for development and look forward to
integrating it in the future.

\subsubsection*{Remaining limitations}

PLINK 2 is designed to meet the needs of tomorrow's genome-wide association
studies and population-genetics research; in both contexts, it is appropriate
to apply a single genomic coordinate system across all samples, and preferred
sample sizes are large enough to make computational efficiency a serious issue.

Whole-exome and whole-genome sequencing also enables detailed study of
structural variations which defy clean representation under a single coordinate
system; and the number of individuals in such studies is typically much smaller
than the tens or even hundreds of thousands which are sometimes required for
effective GWAS.  There are no plans to make PLINK suitable for this type of
analysis; we strongly recommend the use of another software package, such as
PLINK/SEQ \cite{plinkseq2012}, which is explicitly designed for it.  This is
why the PLINK 2 file format will still be substantially less expressive than
VCF.

An important consequence is that, despite its ability to import and export VCF
files, PLINK should not be used for management of genomic data which will be
subject to both types of analysis, because it discards all information which is
not relevant for its preferred type.  However, we will continue to extend
PLINK's ability to interpret VCF-like formats and interoperate with other
popular software.

\section*{Availability and requirements}

\begin{itemize}
\item Project name: PLINK 2
\item Project (source code) home page:
  \url{https://www.cog-genomics.org/plink2/}
  (\url{https://github.com/chrchang/plink-ng})
\item Operating systems: Linux (32/64-bit), OS X (64-bit Intel), Windows
  (32/64-bit)
\item Programming language: C, C++
\item Other requirements (when recompiling): GCC version 4, a few functions
  also require LAPACK 3.2
\item License: GNU General Public License version 3.0 (GPLv3)
\item Any restrictions to use by non-academics: none
\end{itemize}

%%%%%%%%%%%%%%%%%%%%%%%%%%%%%%%%%%%%%%%%%%%%%%
%%                                          %%
%% Backmatter begins here                   %%
%%                                          %%
%%%%%%%%%%%%%%%%%%%%%%%%%%%%%%%%%%%%%%%%%%%%%%

\begin{backmatter}

\section*{Competing interests}
  The authors declare that they have no competing interests.

\section*{Author's contributions}
  SMP and Ch C designed the software.  Ch C drafted the manuscript and did most
of the v1.9 C/C++ programming.  Ca C, SV, and JJL drove early v1.9 feature
development and wrote MATLAB prototype code.  Ca C, LCAMT, SV, SMP, and JJL
assisted with v1.9 software testing.  All authors read and approved the final
manuscript.

\section*{Acknowledgements}
  We thank Stephen D.H. Hsu for helpful discussions.  We also
continue to be thankful to PLINK 1.9 users who perform additional testing of
the program, report bugs, and make useful suggestions.

  Christopher Chang was supported by BGI Hong Kong.  Carson Chow and Shashaank
Vattikuti were supported by the Intramural Research Program of the
NIH, NIDDK.
%%%%%%%%%%%%%%%%%%%%%%%%%%%%%%%%%%%%%%%%%%%%%%%%%%%%%%%%%%%%%
%%                  The Bibliography                       %%
%%                                                         %%
%%  Bmc_mathpys.bst  will be used to                       %%
%%  create a .BBL file for submission.                     %%
%%  After submission of the .TEX file,                     %%
%%  you will be prompted to submit your .BBL file.         %%
%%                                                         %%
%%                                                         %%
%%  Note that the displayed Bibliography will not          %%
%%  necessarily be rendered by Latex exactly as specified  %%
%%  in the online Instructions for Authors.                %%
%%                                                         %%
%%%%%%%%%%%%%%%%%%%%%%%%%%%%%%%%%%%%%%%%%%%%%%%%%%%%%%%%%%%%%

% if your bibliography is in bibtex format, use those commands:
\bibliographystyle{bmc-mathphys} % Style BST file
\bibliography{plink2}      % Bibliography file (usually '*.bib' )

% or include bibliography directly:
% \begin{thebibliography}
% \bibitem{b1}
% \end{thebibliography}

%%%%%%%%%%%%%%%%%%%%%%%%%%%%%%%%%%%
%%                               %%
%% Figures                       %%
%%                               %%
%% NB: this is for captions and  %%
%% Titles. All graphics must be  %%
%% submitted separately and NOT  %%
%% included in the Tex document  %%
%%                               %%
%%%%%%%%%%%%%%%%%%%%%%%%%%%%%%%%%%%

\section*{Figures}
  \begin{figure}[h!]
  \includegraphics[width=0.8\textwidth]{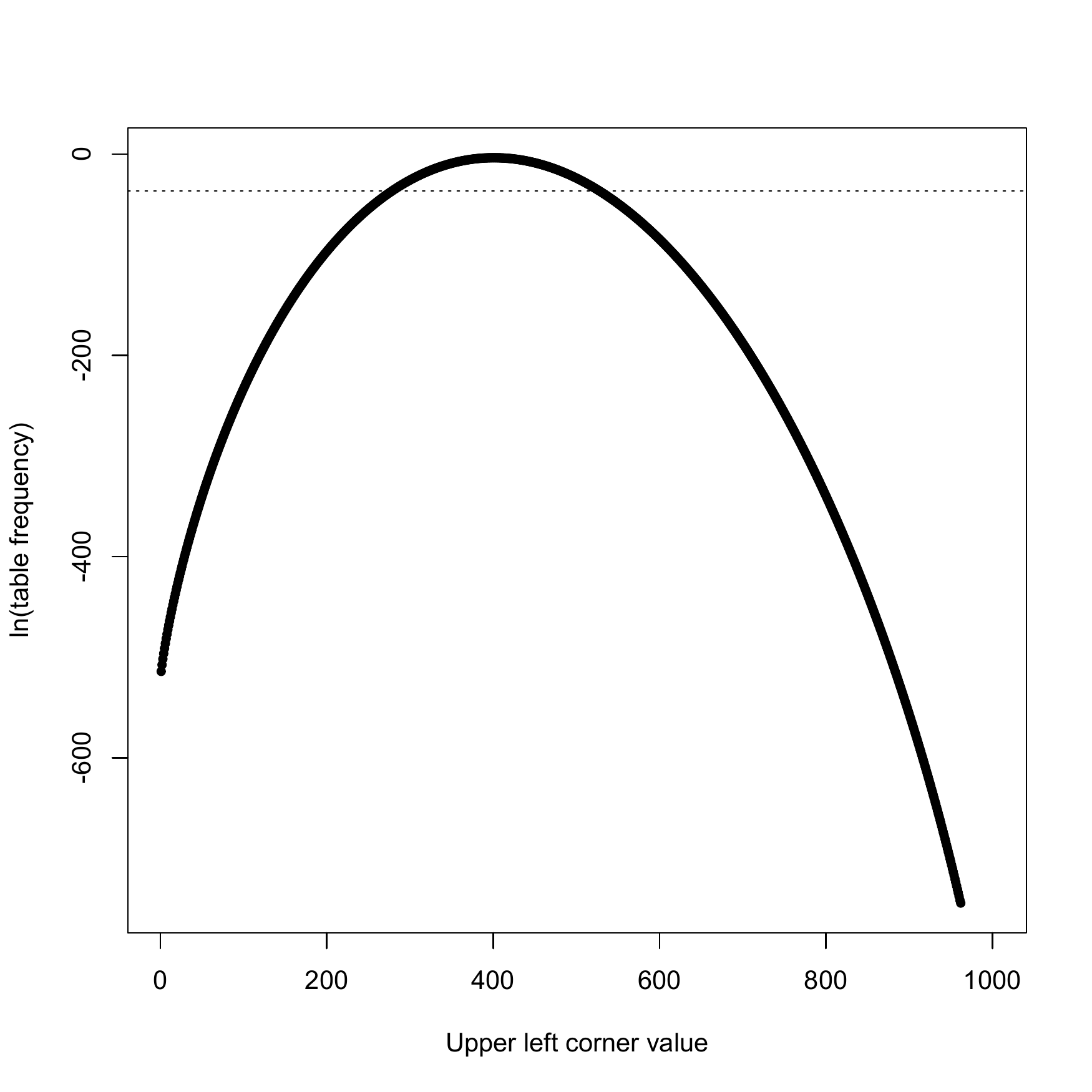}
  \caption{\csentence{2x2 contingency table log-frequencies.}
      This is a plot of relative frequencies of 2x2 contingency tables with top
      row sum 1000, left column sum 40000, and grand total 100000, reflecting a
      low-MAF variant where the difference between the chi-square test and
      Fisher's exact test is relevant.  All such tables with upper left value
      smaller than 278, or larger than 526, have frequency smaller than
      $2^{-53}$ (dotted horizontal line); thus, if the obvious summation
      algorithm is used, they have no impact on the p-value denominator due to
      numerical underflow.  (It can be proven that this underflow has
      negligible impact on accuracy, due to how rapidly the frequencies decay.)
      A few more tables need to be considered when evaluating the numerator,
      but we can usually skip at least 70\%, and this fraction improves as
      problem size increases.}
      \end{figure}

\begin{figure}[h!]
  \includegraphics[width=0.8\textwidth]{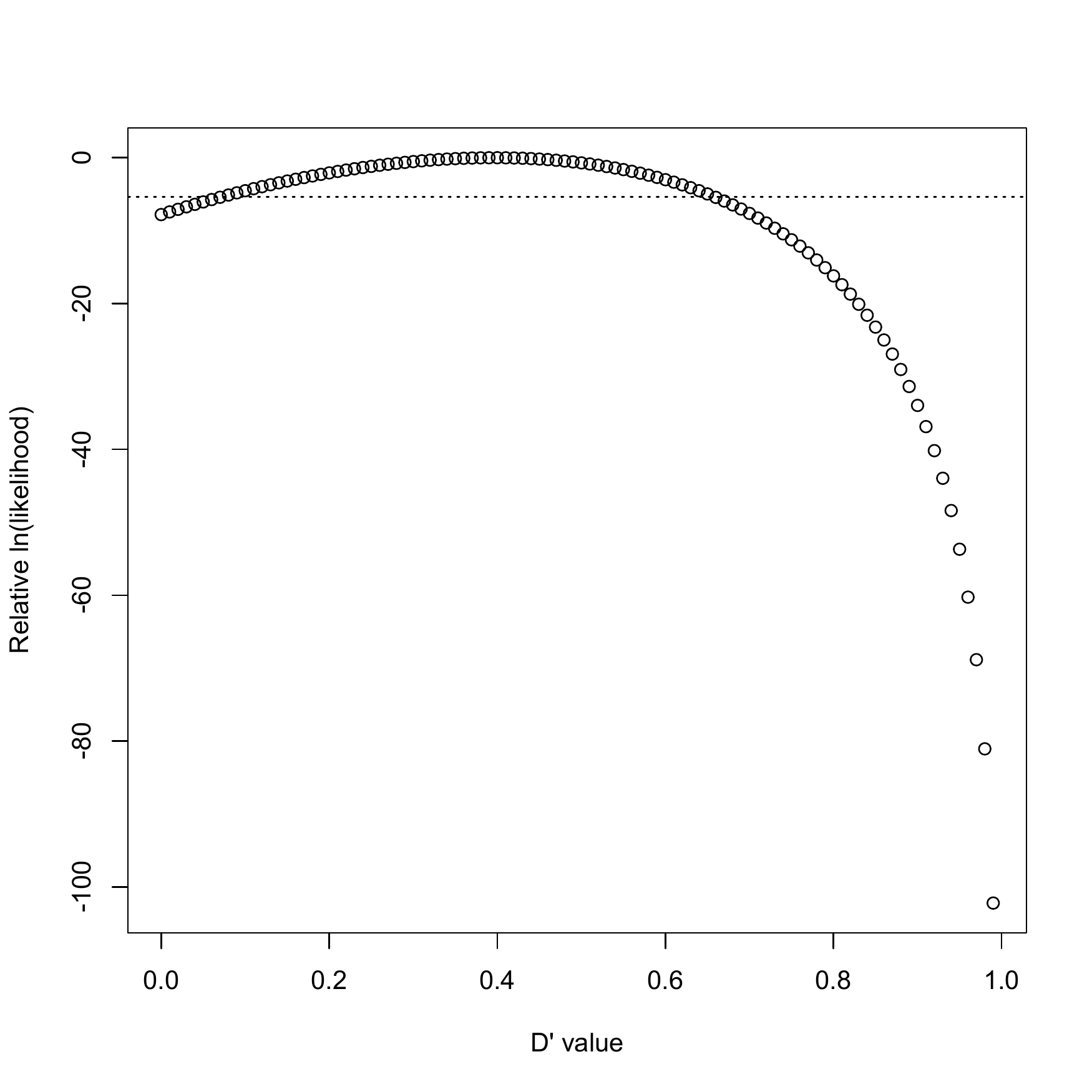}
  \caption{\csentence{Rapid classification of ``recombination'' variant pairs.}
      This is a plot of 101 equally spaced D' log-likelihoods for (rs58108140,
      rs140337953) in 1000 Genomes phase 1, used in Gabriel et al.'s method of
      identifying haplotype blocks.  Whenever the upper end of the 90\%
      confidence interval is smaller than 0.90 (i.e. the rightmost 11
      likelihoods sum to less than 5\% of the total), we have strong evidence
      for historical recombination between the two variants.  After determining
      that $L(D'=x)$ has only one extreme value in [0, 1] and that it's between
      0.39 and 0.40, confirming $L(D'=0.90) < L(D'=0.40)/220$ is enough to
      finish classifying the variant pair (due to monotonicity:
      $L(D'=0.90) \geq L(D'=0.91) \geq \ldots \geq L(D'=1.00)$); evaluation of
      the other 99 likelihoods is now skipped in this case.  The dotted
      horizontal line is at $L(D'=0.40)/220$.}
      \end{figure}

%%
%% Do not use \listoffigures as most will included as separate files

%%%%%%%%%%%%%%%%%%%%%%%%%%%%%%%%%%%
%%                               %%
%% Tables                        %%
%%                               %%
%%%%%%%%%%%%%%%%%%%%%%%%%%%%%%%%%%%
\section*{Tables}
\begin{table}[h!]
\caption{Initialization and basic I/O (\texttt{--freq}).}
\begin{tabular}{c|c|rrr}
  Dataset & Machine & PLINK 1.07 & \textbf{PLINK 1.90} & Ratio \\
  \hline
  \multirow{6}{*}{synth1} & Mac-2 & 7.3 & \textbf{0.24} & 30\phantom{.00} \\
  & Mac-12 & 6.2 & \textbf{0.18} & 34\phantom{.00} \\
  & Linux32-2 & 13.1 & \textbf{0.56} & 23\phantom{.00} \\
  & Linux32-8 & 4.3 & \textbf{0.18} & 24\phantom{.00} \\
  & Linux64-512 & 5.4 & \textbf{0.18} & 27\phantom{.00} \\
  & Win32-2 & 14.3 & \textbf{0.68} & 21\phantom{.00} \\
  & Win64-2 & 9.6 & \textbf{0.33} & 29\phantom{.00} \\
  \hline
  \multirow{6}{*}{synth2} & Mac-2 & 43.3 & \textbf{0.84} & 52\phantom{.00} \\
  & Mac-12 & 38.2 & \textbf{0.34} & 110\phantom{.00} \\
  & Linux32-2 & 80.1 & \textbf{1.9\phantom{0}} & 42\phantom{.00} \\
  & Linux32-8 & 25.2 & \textbf{0.53} & 48\phantom{.00} \\
  & Linux64-512 & 34.1 & \textbf{0.40} & 85\phantom{.00} \\
  & Win32-2 & 83.6 & \textbf{1.3\phantom{0}} & 64\phantom{.00} \\
  & Win64-2 & 70.8 & \textbf{0.55} & 130\phantom{.00} \\
  \hline
  \multirow{6}{*}{chr1snp} & Mac-2 & 52.5 & \textbf{3.5\phantom{0}} & 15\phantom{.00} \\
  & Mac-12 & 40.5 & \textbf{1.3\phantom{0}} & 31\phantom{.00} \\
  & Linux32-2 & 72.9 & \textbf{10.2\phantom{0}} & 7.15 \\
  & Linux32-8 & 29.7 & \textbf{1.4\phantom{0}} & 21\phantom{.00} \\
  & Linux64-512 & 36.8 & \textbf{1.4\phantom{0}} & 26\phantom{.00} \\
  & Win32-2 & 104.3 & \textbf{4.5\phantom{0}} & 23\phantom{.00} \\
  & Win64-2 & 76.8 & \textbf{2.2\phantom{0}} & 35\phantom{.00} \\
  \hline
  \multirow{6}{*}{chr1} & Mac-2 & 403.9 & \textbf{35.0\phantom{0}} & 11.5\phantom{0} \\
  & Mac-12 & 163.9 & \textbf{5.3\phantom{0}} & 31\phantom{.00} \\
  & Linux32-2 & nomem & \textbf{65.3\phantom{0}} & \\
  & Linux32-8 & 134.1 & \textbf{12.8\phantom{0}} & 10.5\phantom{0} \\
  & Linux64-512 & 144.7 & \textbf{5.4\phantom{0}} & 27\phantom{.00} \\
  & Win32-2 & 389.2 & \textbf{21.4\phantom{0}} & 18.2\phantom{0} \\
  & Win64-2 & 285.3 & \textbf{8.1\phantom{0}} & 35\phantom{.00} \\
\end{tabular}
\end{table}

\begin{table}[h!]
\caption{Identity-by-state and complete linkage clustering times.}
\begin{tabular}{c|c|c|rrr}
  Calculation & Dataset & Machine & PLINK 1.07 & \textbf{PLINK 1.90} & Ratio \\
  \hline
  \multirow{18}{*}{IBS matrix only} & \multirow{6}{*}{synth1p} & Mac-2 & 2233.6 & \textbf{1.9} & 1.2k\phantom{0.0} \\
  & & Mac-12 & 1320.4 & \textbf{1.2} & 1.1k\phantom{0.0} \\
  & & Linux32-8 & 1937.2 & \textbf{2.8} & 690\phantom{.0} \\
  & & Linux64-512 & 1492\phantom{.0} & \textbf{3.7} & 400\phantom{.0} \\
  & & Win32-2 & 3219.0 & \textbf{7.2} & 450\phantom{.0} \\
  & & Win64-2 & 2674.4 & \textbf{1.5} & 1.8k\phantom{0.0} \\ \cline{2-6}
  & \multirow{6}{*}{synth2p} & Mac-2 & $\sim$190k\phantom{00.0} & \textbf{118.8} & 1.6k\phantom{0.0} \\
  & & Mac-12 & $\sim$99k\phantom{00.0} & \textbf{23.5} & 4.2k\phantom{0.0} \\
  & & Linux32-8 & 152.5k\phantom{00} & \textbf{214.3} & 710\phantom{.0} \\
  & & Linux64-512 & $\sim$98k\phantom{00.0} & \textbf{25.3} & 3.9k\phantom{0.0} \\
  & & Win32-2 & $\sim$270k\phantom{00.0} & \textbf{654.5} & 410\phantom{.0} \\
  & & Win64-2 & $\sim$200k\phantom{00.0} & \textbf{104.6} & 1.9k\phantom{0.0} \\ \cline{2-6}
  & \multirow{6}{*}{chr1snp} & Mac-2 & $\sim$26k\phantom{00.0} & \textbf{17.5} & 1.5k\phantom{0.0} \\
  & & Mac-12 & 13.4k\phantom{00} & \textbf{12.6} & 1.06k\phantom{.0} \\
  & & Linux32-8 & 18.4k\phantom{00} & \textbf{30.9} & 600\phantom{.0} \\
  & & Linux64-512 & $\sim$14k\phantom{00.0} & \textbf{43.1} & 320\phantom{.0} \\
  & & Win32-2 & 32.7k\phantom{00} & \textbf{95.9} & 341\phantom{.0} \\
  & & Win64-2 & $\sim$26k\phantom{00.0} & \textbf{15.3} & 1.7k\phantom{0.0} \\
  \hline
  \multirow{18}{*}{Basic clustering} & \multirow{6}{*}{synth1p} & Mac-2 & 2315.7 & \textbf{2.7} & 860\phantom{.0} \\
  & & Mac-12 & 1317.9 & \textbf{2.0} & 660\phantom{.0} \\
  & & Linux32-8 & 1898.7 & \textbf{4.1} & 460\phantom{.0} \\
  & & Linux64-512 & 1496\phantom{.0} & \textbf{4.5} & 330\phantom{.0} \\
  & & Win32-2 & 3301.7 & \textbf{9.1} & 360\phantom{.0} \\
  & & Win64-2 & 2724.5 & \textbf{1.9} & 1.4k\phantom{0.0} \\ \cline{2-6}
  & \multirow{6}{*}{synth2p} & Mac-2 & $\sim$230k\phantom{00.0} & \textbf{245.6} & 940\phantom{.0} \\
  & & Mac-12 & $\sim$140k\phantom{00.0} & \textbf{123.9} & 1.1k\phantom{0.0} \\
  & & Linux32-8 & 197.1k\phantom{00} & \textbf{395.6} & 498\phantom{.0} \\
  & & Linux64-512 & $\sim$125k\phantom{00.0} & \textbf{143.3} & 872\phantom{.0} \\
  & & Win32-2 & $\sim$440k\phantom{00.0} & \textbf{976.7} & 450\phantom{.0} \\
  & & Win64-2 & $\sim$270k\phantom{00.0} & \textbf{127.9} & 2.1k\phantom{0.0} \\ \cline{2-6}
  & \multirow{6}{*}{chr1snp} & Mac-2 & $\sim$26k\phantom{00.0} & \textbf{18.4} & 1.4k\phantom{0.0} \\
  & & Mac-12 & 13.6k\phantom{00} & \textbf{13.5} & 1.01k\phantom{.0} \\
  & & Linux32-8 & 18.5k\phantom{00} & \textbf{33.4} & 554\phantom{.0} \\
  & & Linux64-512 & $\sim$14k\phantom{00.0} & \textbf{44.2} & 320\phantom{.0} \\
  & & Win32-2 & 33.2k\phantom{00} & \textbf{95.0} & 349\phantom{.0} \\
  & & Win64-2 & $\sim$26k\phantom{00.0} & \textbf{15.8} & 1.6k\phantom{0.0} \\
  \hline
  \multirow{18}{*}{IBD report} & \multirow{6}{*}{synth1p} & Mac-2 & 2230.1 & \textbf{12.4} & 180\phantom{.0} \\
  & & Mac-12 & 1346.2 & \textbf{2.4} & 560\phantom{.0} \\
  & & Linux32-8 & 2019.9 & \textbf{12.4} & 163\phantom{.0} \\
  & & Linux64-512 & 1494\phantom{.0} & \textbf{5.0} & 300\phantom{.0} \\
  & & Win32-2 & 3446.3 & \textbf{42.2} & 81.7 \\
  & & Win64-2 & 2669.8 & \textbf{15.1} & 177\phantom{.0} \\ \cline{2-6}
  & \multirow{6}{*}{synth2p} & Mac-2 & $\sim$190k\phantom{00.0} & \textbf{447.1} & 420\phantom{.0} \\
  & & Mac-12 & $\sim$99k\phantom{00.0} & \textbf{50.3} & 2.0k\phantom{0.0} \\
  & & Linux32-8 & 161.4k\phantom{00} & \textbf{618.7} & 261\phantom{.0} \\
  & & Linux64-512 & $\sim$98k\phantom{00.0} & \textbf{57.4} & 1.7k\phantom{0.0} \\
  & & Win32-2 & $\sim$270k\phantom{00.0} & \textbf{1801.1} & 150\phantom{.0} \\
  & & Win64-2 & $\sim$200k\phantom{00.0} & \textbf{541.0} & 370\phantom{.0} \\ \cline{2-6}
  & \multirow{6}{*}{chr1snp} & Mac-2 & $\sim$26k\phantom{00.0} & \textbf{24.8} & 1.0k\phantom{0.0} \\
  & & Mac-12 & 13.4k\phantom{00} & \textbf{14.6} & 918\phantom{.0} \\
  & & Linux32-8 & 18.5k\phantom{00} & \textbf{53.5} & 346\phantom{.0} \\
  & & Linux64-512 & $\sim$14k\phantom{00.0} & \textbf{46.5} & 300\phantom{.0} \\
  & & Win32-2 & 33.1k\phantom{00} & \textbf{199.2} & 166\phantom{.0} \\
  & & Win64-2 & $\sim$26k\phantom{00.0} & \textbf{25.1} & 1.0k\phantom{0.0} \\
\end{tabular}
\end{table}

\begin{table}[h!]
\caption{Genomic relationship matrix calculation times.}
\begin{tabular}{c|c|rrr}
  Dataset & Machine & GCTA & \textbf{PLINK 1.90} & Ratio \\
  \hline
  \multirow{6}{*}{synth1p} & Mac-2 & 222.2 & \textbf{7.2} & 31\phantom{.00} \\
  & Mac-12 & 184.7 & \textbf{5.0} & 37\phantom{.00} \\
  & Linux32-8 & 248.4 & \textbf{10.9} & 22.8\phantom{0} \\
  & Linux64-512 & 4.4 & \textbf{8.3} & 0.53 \\
  & Win32-2 & 373.1 & \textbf{39.3} & 9.5\phantom{0} \\
  & Win64-2 & 367.2 & \textbf{6.6} & 56\phantom{.00} \\
  \hline
  \multirow{6}{*}{synth2p} & Mac-2 & nomem & \textbf{805.8} & \\
  & Mac-12 & 17.0k\phantom{00} & \textbf{138.3} & 123\phantom{.00} \\
  & Linux32-8 & nomem & \textbf{1153.4} & \\
  & Linux64-512 & 65.1 & \textbf{166.0} & 0.39 \\
  & Win32-2 & nomem & \textbf{2007.2} & \\
  & Win64-2 & nomem & \textbf{450.1} & \\
  \hline
  \multirow{6}{*}{chr1snp} & Mac-2 & nomem & \textbf{87.1} & \\
  & Mac-12 & 2260.9 & \textbf{50.9} & 44.4\phantom{0} \\
  & Linux32-8 & nomem & \textbf{94.3} & \\
  & Linux64-512 & 58.3 & \textbf{86.9} & 0.67 \\
  & Win32-2 & nomem & \textbf{317.5} & \\
  & Win64-2 & nomem & \textbf{65.7} & \\
\end{tabular}
\end{table}

\begin{table}[h!]
\caption{\texttt{--indep-pairwise} runtimes.}
\begin{tabular}{c|c|c|rrr}
  Parameters & Dataset & Machine & PLINK 1.07 & \textbf{PLINK 1.90} & Ratio \\
  \hline
  \multirow{6}{*}{\texttt{50 5 0.5}} & \multirow{6}{*}{synth1} & Mac-2 & 701.3 & \textbf{0.63} & 1.1k\phantom{0} \\
  & & Mac-12 & 569.4 & \textbf{0.55} & 1.0k\phantom{0} \\
  & & Linux32-8 & 572.7 & \textbf{0.95} & 600 \\
  & & Linux64-512 & 462\phantom{.0} & \textbf{0.60} & 770 \\
  & & Win32-2 & 1163.9 & \textbf{3.2\phantom{0}} & 360 \\
  & & Win64-2 & 1091.9 & \textbf{1.0\phantom{0}} & 1.1k\phantom{0} \\
  \hline
  \multirow{6}{*}{\texttt{700 70 0.7}} & \multirow{6}{*}{synth2} & Mac-2 & $\sim$120k\phantom{00.0} & \textbf{31.9\phantom{0}} & 3.8k\phantom{0} \\
  & & Mac-12 & 63.0k\phantom{00} & \textbf{20.6\phantom{0}} & 3.06k \\
  & & Linux32-8 & 57.4k\phantom{00} & \textbf{66.0\phantom{0}} & 870 \\
  & & Linux64-512 & $\sim$120k\phantom{00.0} & \textbf{26.4\phantom{0}} & 4.5k\phantom{0} \\
  & & Win32-2 & 139.3k\phantom{00} & \textbf{127.3\phantom{0}} & 1.09k \\
  & & Win64-2 & $\sim$200k\phantom{00.0} & \textbf{22.9\phantom{0}} & 8.7k\phantom{0} \\
  \hline
  \multirow{12}{*}{\texttt{20000 2000 0.5}} & \multirow{6}{*}{chr1} & Mac-2 & nomem & \textbf{1520.1\phantom{0}} & \\
  & & Mac-12 & nomem & \textbf{1121.7\phantom{0}} & \\
  & & Linux32-8 & nomem & \textbf{4273.9\phantom{0}} & \\
  & & Linux64-512 & $\sim$950k\phantom{00.0} & \textbf{1553.3\phantom{0}} & 610 \\
  & & Win32-2 & nomem & \textbf{4912.7\phantom{0}} & \\
  & & Win64-2 & nomem & \textbf{1205.1\phantom{0}} & \\ \cline{2-6}
  & \multirow{6}{*}{1000g} & Mac-2 & nomem & \textbf{20.5k\phantom{000}} & \\
  & & Mac-12 & nomem & \textbf{14.5k\phantom{000}} & \\
  & & Linux32-8 & nomem & \textbf{54.5k\phantom{000}} & \\
  & & Linux64-512 & $\sim$13000k\phantom{00.0} & \textbf{20.2k\phantom{000}} & 640 \\
  & & Win32-2 & nomem & \textbf{64.5k\phantom{000}} & \\
  & & Win64-2 & nomem & \textbf{14.7k\phantom{000}} & \\
\end{tabular}
\end{table}

\begin{table}[h!]
\caption{\texttt{--blocks} runtimes.}
\begin{tabular}{c|c|c|rrr}
  Parameters & Dataset & Machine & Haploview 4.2 & PLINK 1.07 & \textbf{PLINK 1.90} \\
  \hline
  \multirow{6}{*}{\texttt{--ld-window-kb 500}} & \multirow{6}{*}{synth1} & Mac-2 & nomem & 3198.4 & \textbf{1.7} \\
  & & Mac-12 & $\sim$45k\phantom{?} & 3873.0 & \textbf{1.3} \\
  & & Linux32-2 & nomem & 5441.1 & \textbf{3.4} \\
  & & Linux64-512 & $\sim$57k\phantom{?} & 2323.4 & \textbf{2.9} \\
  & & Win32-2 & nomem & 9803.4 & \textbf{8.9} \\
  & & Win64-2 & $\sim$51k\phantom{?} & 5513.4 & \textbf{2.8} \\
  \hline
  \multirow{6}{*}{\texttt{--ld-window-kb 1000}} & \multirow{6}{*}{synth1} & Mac-2 & nomem & 6185.7 & \textbf{2.2} \\
  & & Mac-12 & $\sim$45k\phantom{?} & 7394.4 & \textbf{9.8} \\
  & & Linux32-2 & nomem & 9876.8 & \textbf{10.0} \\
  & & Linux64-512 & $\sim$57k\phantom{?} & 4462.1 & \textbf{3.9} \\
  & & Win32-2 & nomem & 18925.7 & \textbf{17.3} \\
  & & Win64-2 & $\sim$51k\phantom{?} & 10.3k\phantom{00} & \textbf{3.6} \\
  \hline
  \multirow{6}{*}{\texttt{--ld-window-kb 500}} & \multirow{6}{*}{chr1} & Mac-2 & nomem & $\sim$2700k?\phantom{0.0} & \textbf{550.9} \\
  & & Mac-12 & nomem & $\sim$3600k?\phantom{0.0} & \textbf{426.0} \\
  & & Linux32-2 & nomem & $\sim$4300k?\phantom{0.0} & \textbf{1288.4} \\
  & & Linux64-512 & $\sim$440k? & $\sim$2600k?\phantom{0.0} & \textbf{1119.7} \\
  & & Win32-2 & nomem & $\sim$17000k?\phantom{0.0} & \textbf{4535.8} \\
  & & Win64-2 & nomem & $\sim$5700k?\phantom{0.0} & \textbf{1037.2} \\
\end{tabular}
\end{table}

\begin{table}[h!]
\caption{Association analysis max(T) permutation test times.
(\texttt{--mperm 10000 --seed 1})}
\begin{tabular}{c|c|c|rrrr}
  Other parameter(s) & Dataset & Machine & PLINK 1.07 & PERMORY 1.1 & \textbf{PLINK 1.90} & Ratio \\
  \hline
  \multirow{12}{*}{\texttt{--trend} (C/C)} & \multirow{6}{*}{synth1} & Mac-2 & $\sim$20k & & \textbf{18.7} & 1.1k\phantom{000} \\
  & & Mac-12 & $\sim$16k & & \textbf{2.8} & 5.7k\phantom{000} \\
  & & Linux32-2 & $\sim$21k & & \textbf{65.0} & 320\phantom{.00} \\
  & & Linux64-512 & $\sim$17k & 285.0 & \textbf{2.8} & \\
  & & Win32-2 & $\sim$35k & 1444.2 & \textbf{61.5} & \\
  & & Win64-2 & $\sim$25k & 889.7 & \textbf{14.4} & \\ \cline{2-7}
  & \multirow{6}{*}{synth2} & Mac-2 & $\sim$170k & & \textbf{42.4} & 4.0k\phantom{000} \\
  & & Mac-12 & $\sim$180k & & \textbf{6.4} & 28k\phantom{00.00} \\
  & & Linux32-2 & $\sim$410k & & \textbf{391.0} & 1.0k\phantom{000} \\
  & & Linux64-512 & $\sim$200k & 580.9 & \textbf{13.7} & \\
  & & Win32-2 & $\sim$1100k & 2362.5 & \textbf{198.0} & \\
  & & Win64-2 & $\sim$370k & 1423.6 & \textbf{34.0} & \\
  \hline
  \multirow{12}{*}{\texttt{--fisher} (C/C)} & \multirow{6}{*}{synth1} & Mac-2 & $\sim$150k & & \textbf{21.9} & 6.9k\phantom{000} \\
  & & Mac-12 & $\sim$150k & & \textbf{3.7} & 41k\phantom{00.00} \\
  & & Linux32-2 & $\sim$170k & & \textbf{57.8} & 2.9k\phantom{000} \\
  & & Linux64-512 & $\sim$120k & & \textbf{3.4} & 35k\phantom{00.00} \\
  & & Win32-2 & $\sim$440k & & \textbf{64.9} & 6.8k\phantom{000} \\
  & & Win64-2 & $\sim$200k & & \textbf{22.0} & 9.1k\phantom{000} \\ \cline{2-7}
  & \multirow{6}{*}{synth2} & Mac-2 & $\sim$890k & & \textbf{49.8} & 18k\phantom{00.00} \\
  & & Mac-12 & $\sim$690k & & \textbf{7.6} & 91k\phantom{00.00} \\
  & & Linux32-2 & $\sim$1300k & & \textbf{393.7} & 3.3k\phantom{000} \\
  & & Linux64-512 & $\sim$720k & & \textbf{13.0} & 55k\phantom{00.00} \\
  & & Win32-2 & $\sim$3600k & & \textbf{208.3} & 17k\phantom{00.00} \\
  & & Win64-2 & $\sim$1700k & & \textbf{35.6} & 48k\phantom{00.00} \\
  \hline
  \multirow{6}{*}{\texttt{--assoc} (QT)} & \multirow{6}{*}{synth1} & Mac-2 & $\sim$30k & & \textbf{148.0} & 200\phantom{.00} \\
  & & Mac-12 & $\sim$22k & & \textbf{22.6} & 970\phantom{.00} \\
  & & Linux32-2 & $\sim$68k & & \textbf{847.2} & 80\phantom{.00} \\
  & & Linux64-512 & $\sim$29k & & \textbf{29.2} & 990\phantom{.00} \\
  & & Win32-2 & $\sim$58k & & \textbf{896.1} & 65\phantom{.00} \\
  & & Win64-2 & $\sim$36k & & \textbf{264.2} & 140\phantom{.00} \\
  \hline
  \multirow{6}{*}{\texttt{--assoc lin} (QT)} & \multirow{6}{*}{synth1} & Mac-2 & & & \textbf{606.8} & \\
  & & Mac-12 & & & \textbf{34.7} & \\
  & & Linux32-2 & & & \textbf{3212.6} & \\
  & & Linux64-512 & & 1259.8 & \textbf{46.4} & 27.2\phantom{0} \\
  & & Win32-2 & & 2115.7 & \textbf{3062.7} & 0.69 \\
  & & Win64-2 & & 972.6 & \textbf{336.6} & 2.89 \\
\end{tabular}
\end{table}

%% Use of \listoftables is discouraged.
%%

%%%%%%%%%%%%%%%%%%%%%%%%%%%%%%%%%%%
%%                               %%
%% Additional Files              %%
%%                               %%
%%%%%%%%%%%%%%%%%%%%%%%%%%%%%%%%%%%

\begin{figure}
  \centering
  \includegraphics[page=1,scale=0.75]{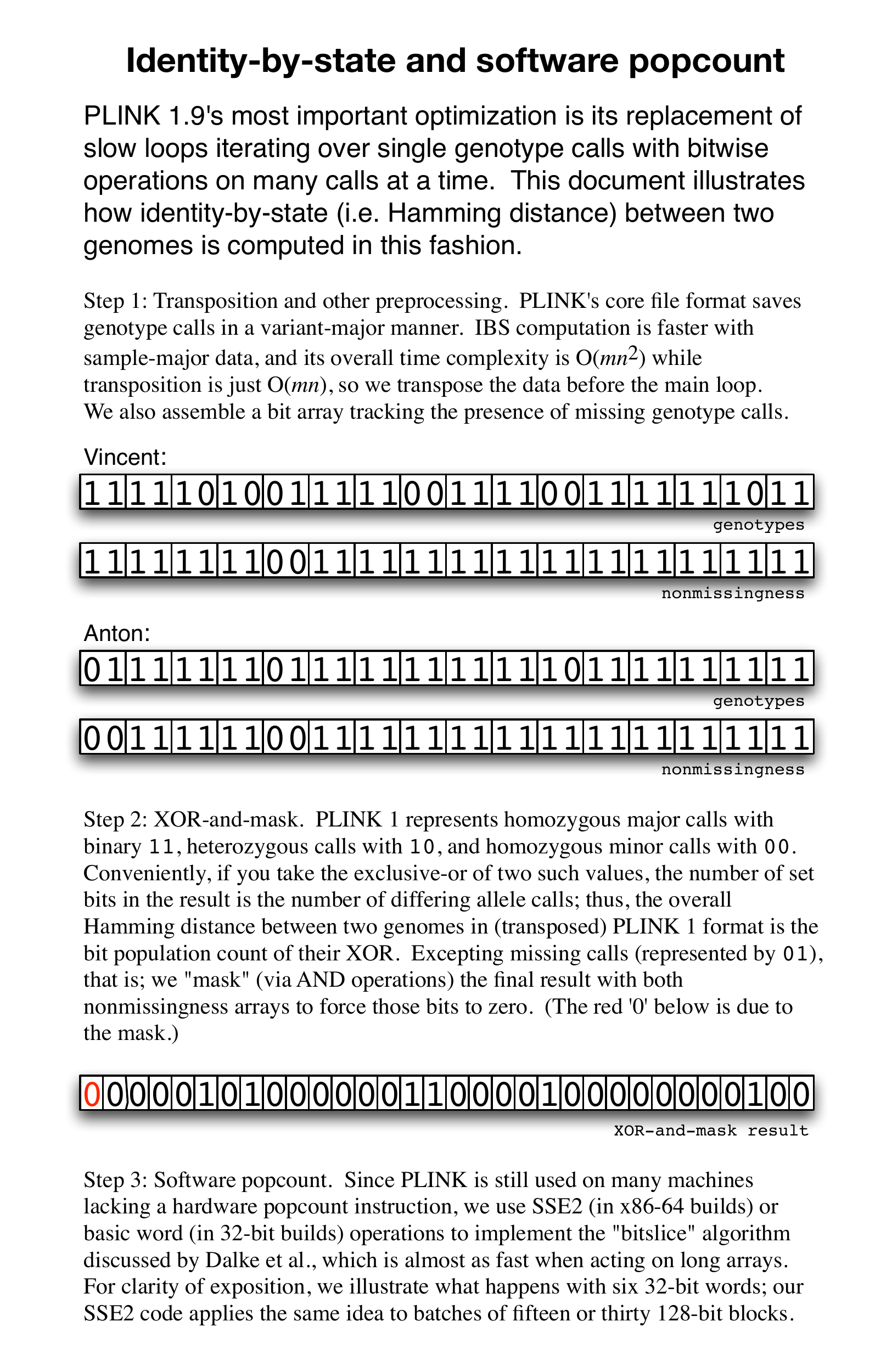}
\end{figure}

\begin{figure}
  \centering
  \includegraphics[page=2,scale=0.75]{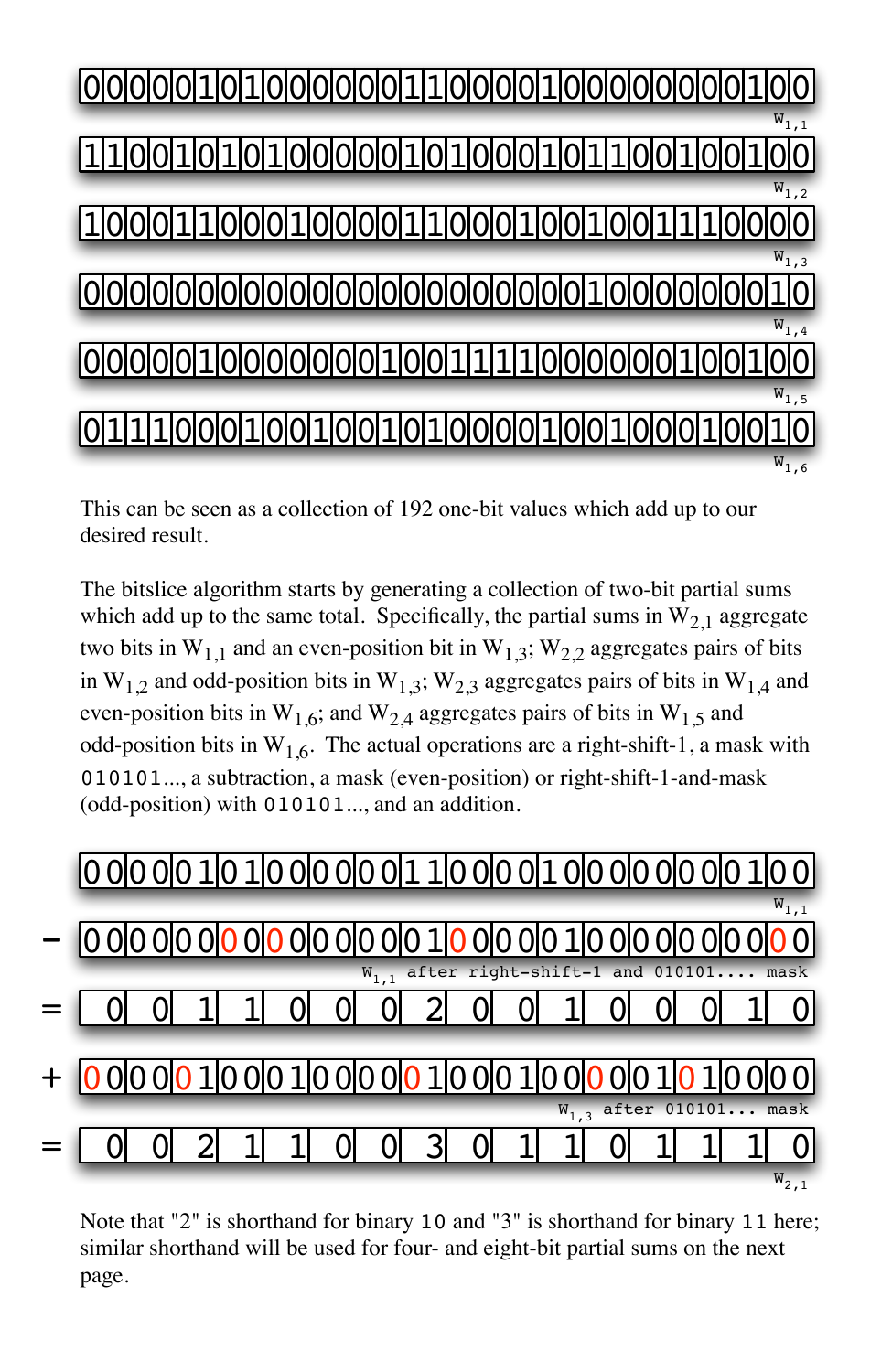}
\end{figure}

\begin{figure}
  \centering
  \includegraphics[page=3,scale=0.75]{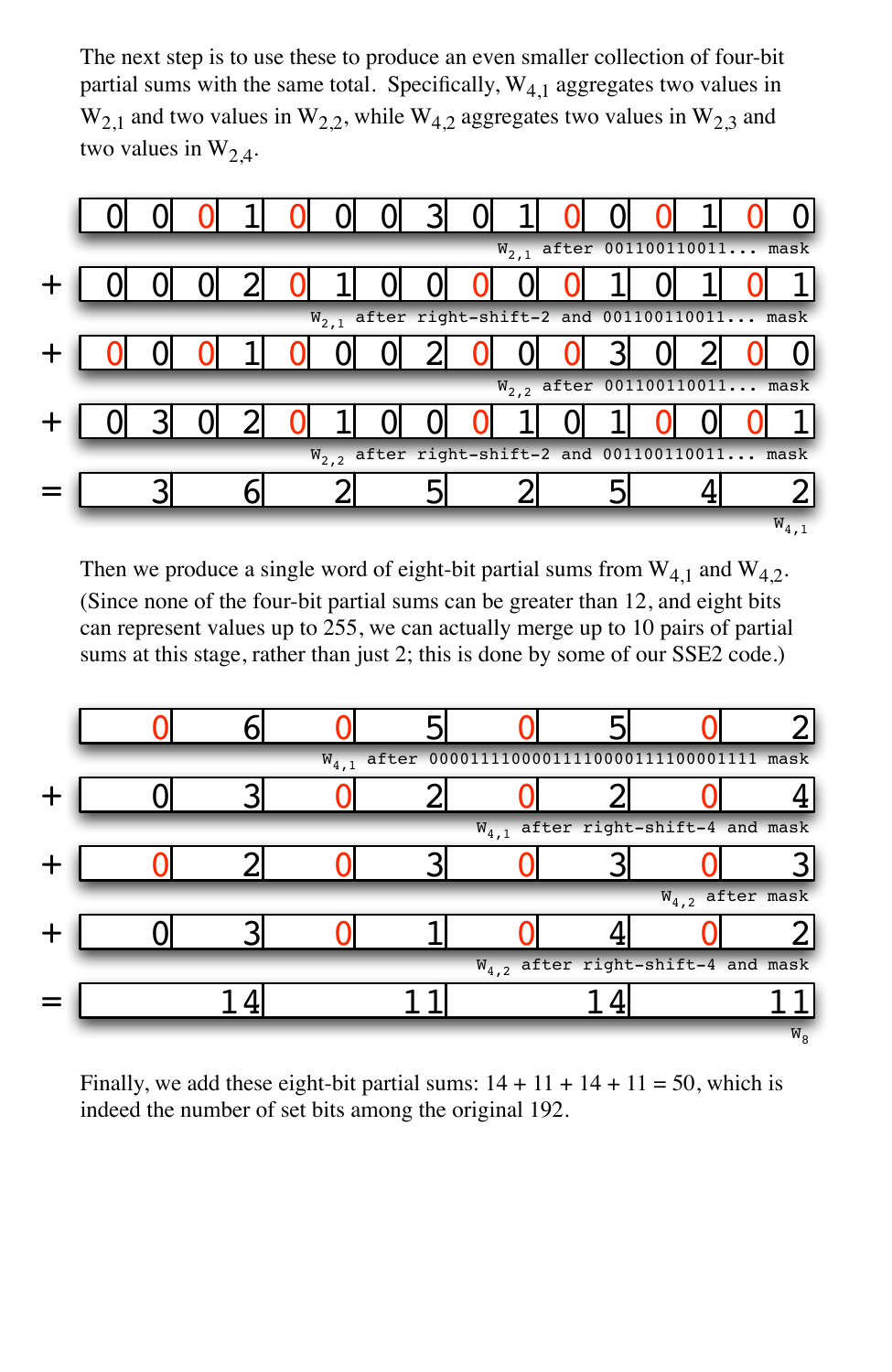}
\end{figure}

\end{backmatter}
\end{document}